\documentclass[12pt,epsf]{article}
\usepackage{epsfig}

\newcommand{\twofigures}[3]{\begin{figure}[htdp]
\centering \leavevmode\epsfxsize=2in\epsfbox{#1.eps}
\leavevmode\epsfxsize=2in\epsfbox{#2.eps} \caption{\small {\em
#3}\label{#1}}
\end{figure}}

\newcommand{\be}{\begin{equation}}
\newcommand{\ee}{\end{equation}}
\newcommand{\bea}{\begin{eqnarray}}
\newcommand{\eea}{\end{eqnarray}}

\begin{document}

\begin{flushright}
SLAC-PUB-10780\\
\end{flushright}

\bigskip\bigskip
\begin{center}
{\bf\large Cosmic Microwave Background Fluctuation Amplitude from
Dark Energy De-Coherence\footnote{\baselineskip=12pt Work
supported by Department of Energy contract DE--AC03--76SF00515.}}
\end{center}

\begin{center}
James V. Lindesay\footnote{Permanent address, Physics Department, Howard University,
Washington, D.C. 20059}, jlslac@slac.stanford.edu\\
H. Pierre Noyes, noyes@slac.stanford.edu \\
Stanford Linear Accelerator Center MS 81,
Stanford University \\
2575 Sand Hill Road, Menlo Park CA 94025\\
\end{center}

{\small Standard cosmology connects the scale generated at late
times by the cosmological constant with that generated in early
times by the energy dynamics.  Assuming dark energy de-coherence
occurs during early times when the scale parameter expansion rate
is no longer supra-luminal specifies a class of cosmological
models in which the cosmic microwave background fluctuation
amplitude at last scattering is approximately $10^{-5}$. }

PACS  numbers: 98.80.Bp, 98.80.Jk, 95.30.Sf

\bigskip
\setcounter{equation}{0}

The luminosities of distant Type Ia
supernovae show that the rate of expansion of the
universe has been accelerating for several
giga-years\cite{TypeIa}. This conclusion is independently
confirmed by analysis of the Cosmic Microwave Background
radiation\cite{PDG,WMAP}.  Both results are in quantitative
agreement with a (positive) cosmological constant fit to the data.
The existence of a cosmological constant / dark energy density
defines a length scale that must be incorporated in any
description of the evolution of our universe.   When the dynamics
of the cosmology is made consistent with this scale, it is expected
that the usual microscopic interactions of relativistic quantum
mechanics (QED, QCD, etc) cannot contribute to cosmological
(gravitational) equilibrations when the Friedmann-Robertson-Walker
(FRW) scale expansion rate is supra-luminal, $\dot{R}>c$.  The
cosmological dark energy density is expected to decouple from the
energy density in the Friedmann-Lemaitre(FL) equations when the
FRW scale expansion is no longer supra-luminal, at
which time the microscopic interactions open new degrees of
freedom.

Prior to de-coherence, the coherence which preserves the
uniform density needed to make the FL
dynamical equations meaningful must be maintained by
\emph{supra-luminal} (eg. gravitational) correlations and not by
the luminal or sub-luminal microscopic exchanges that are available
after de-coherence. In this sense the energy which de-coheres
\emph{must be ``dark"}.  In what follows it is called \emph{dark
energy}. If the usual backward extrapolations from the present to
\emph{de-coherence time} are accepted, it is expected that the decoupling
of the evolution of the dark energy from that of
the microscopic degrees of freedom will occur during a time when the
mass density which drives the FL dynamical equations is dominated
by radiation, and at a finite time \emph{earlier} than big-bang
nucleosynthesis.  The basic assumption in this paper is that the
dark energy which is fixed during
de-coherence is to be identified with the cosmological constant.
It is shown that the expected amplitude of fluctuations driven by the
dark energy de-coherence process is of the order needed to evolve
into the fluctuations observed in cosmic microwave background
radiation and in galactic clustering. From the time of dark energy
de-coherence until the scale factor expansion rate again becomes
supra-luminal due to the cosmological
constant, the usual expansion rate evolution predicted by the FL
dynamics is expected to hold.

The horizon problem arises from
the large scale homogeneity and isotropy of 
luminally disconnected regions of the observed universe.  
The present age of the universe can be estimated from
the Hubble rate using $H_o t_o \cong {2 \over 3 \sqrt{\Omega_{\Lambda o}}} log
\left ( 1+\sqrt{\Omega_{\Lambda o}} \over \sqrt{1-\Omega_{\Lambda o}}   \right ) 
\cong 0.96 $ to be around 13 Gyr
(see reference \cite{PDG}, page 193, Eq. 19.30).
Here $\Omega_{\Lambda o}$ is
the present normalized dark energy density
for a spatially flat cosmology.
If the size of the observable universe
today is taken to be of the order of the Hubble scale ${c \over
H_o} \approx 10^{28}cm$,
then its size at the Planck scale would have been of the order
$\sim 10^{-4}cm$. Since the Planck length is of the
order $L_P \sim 10^{-33}cm$, then there are expected to be
about $(10^{29})^3 \sim 10^{87}$ (luminal) causally disconnected regions in the
sky.  Further, examining the ratio of the present conformal time
$\eta_o$ with that during recombination ${\eta_o \over \eta_*}
\sim 100$, the subsequent expansion is expected to imply that
light from the cosmic microwave background would come from about
$100^3=10^6$ disconnected regions.  Yet, angular correlations of
the fluctuations have been accurately measured by several
experiments\cite{WMAP}.
This homogeneity and correlation is indicative of some
form of cosmological coherence in earlier times.

The general approach used here is to start from well
understood macrophysics,
assume that the physics of de-coherence defines a
cosmological scale parameter, and end the 
examination of the backward extrapolation of cosmological physics at the
time when the rate of expansion of that scale parameter is
the velocity of light. For times after that transition
there is general confidence that well
understood micro- \emph{and} macro-physics are valid
at the cosmological level.
The process that takes the cosmology from the very early universe
(i.e. prior to de-coherence) through this boundary
will be called {\it gravitational dark energy de-coherence}. 
The FRW scale parameter is used
to compare cosmological scales to those
scales relevant for microscopic physics, which define the lengths of rulers,
ticks of clocks, mass of particles, and temperatures of
thermodynamic systems. The calculations presented here will not
depend on the present horizon scale, which is an accident of
history.  Global gravitational
coherence prior to de-coherence (i.e. the assumption that the FL
equations still apply in the very early universe) solves the
horizon problem because the gravitational correlations implied by
the FL equations are supra-luminal; it is hypothesized that the same
will be true of any type of dark energy to be considered. A preliminary
discussion of why \emph{quantum coherence} in the very early
universe might also provide the requisite dark energy scale has been
presented elsewhere\cite{ANPA}.

The existence of a cosmological constant $\Lambda$
introduces a natural length scale for the late time FRW metric.  
If we express the FRW scale parameter $R(t)$
which results from the physics of de-coherence
in units consistent with those of 
the dark energy density $\rho_\Lambda={\Lambda c^4 \over 8 \pi G_N}$ 
and the FL energy density $\rho$ which drive
the expansion in the FL equations, then the expansion rate $\dot{R}$ is expected to have physical significance
with respect to the cosmological dark energy.  In particular, for very late times, the De Sitter horizon
$R_{DeSitter}=\sqrt{{3 \over \Lambda}} \equiv R_\Lambda \approx 1.5 \times 10^{28} cm \approx 1.6 \times 10^{10} ly$
scales the exponential expansion rate of the scale parameter.
For regions beyond this horizon, no luminal signals can reach $r=0$, and information from those inaccessible
regions is limited to what can be ascertained from the temperature of this horizon
$k_B T_{DeSitter}={\hbar c \over 2 \pi R_{DeSitter}} \sim 2 \times 10^{-30} \: \raisebox{1ex}{\scriptsize o} K$.
For a cosmology primarily driven by a positive cosmological constant,
\emph{any} FRW scale parameter has a value equal to the De Sitter horizon scale
when the late time expansion rate is that of light, $\dot{R}=c$.
An understanding of the physics of de-coherence allows one to
use the known value of the cosmological constant (dark energy) to determine the
behavior of the scale parameter during early times.
It is therefore meaningful to use this scale parameter to describe
the subsequent dynamics of the cosmology.
If this is the scale parameter to be used in the FL equations, an ``open"
cosmology is excluded, since (as will be shown) at no time in such a
cosmology is $\dot{R} \leq c$.  Likewise, a ``closed"
cosmology can be shown to never expand to an extent that would allow structure formation\cite{ANPA}.
The only cosmology consistent with the observed phenomenology is then exactly spatially flat.
In the development that follows, the evolution of a spatially flat cosmology
using the natural scale factor determined by the physics of dark energy de-coherence
will be examined.

Prior to the de-coherence scale condition
$\dot{R}_{DC} =c$, gravitational influences on scales $R<R_{DC}$
must propagate (at least) at the rate of the gravitational scale expansion, and
microscopic interactions (which are assumed to propagate no faster than c)
cannot contribute to cosmological scale equilibrations.  
If the expansion rate is supra-luminal $\dot{R}>c$,
scattering states cannot form decomposed (de-coherent)
clusters of the type described in references \cite{AKLN,LMNP}
on cosmological scales, i.e. incoherent decomposed
clusters cannot be cosmologically formulated.  Since
the development of the large number statistics needed to define a
cosmologically significant temperature requires an equilibration of interacting
``microstates", any mechanism for the cosmological redistribution
of those microstates on time scales more rapid than the cosmological
expansion rate can only be through gravitational interactions.

The question now arises as to whether dark energy is geometric or
dynamical in origin.  From the form of Einstein's equation
\be
\mathcal{R}_{\mu \nu} - {1 \over 2} g_{\mu \nu} \mathcal{R} \: = \:
{8 \pi G_N \over c^4} T_{\mu \nu} + \Lambda g_{\mu \nu} ,
\ee
if the term involving the cosmological constant should most naturally
appear on the left hand side of the equation, dark energy would be considered geometric in origin.  
If this term is geometric in origin, one would expect $\Lambda$
to be a fundamental constant which scales with the FRW/FL cosmology consistently
with the vanishing divergence of the Einstein tensor.
However, if it arises from the process of de-coherence at $\dot{R}=c$,
this means that it is fixed by a physical condition being met,
and thus would not be a purely geometric constant.

During de-coherence, the Friedmann-Lemaitre(FL) equations
which relate the rate and acceleration of the expansion to the densities,
are given by
\be
H^2 (t) \: = \: \left ( {\dot{R} \over R} \right ) ^2 \: = \: {8 \pi G_N \over 3 c^2}
\left ( \rho + \rho_\Lambda  \right ) \, - \, {k c^2 \over R^2}
\label{Hubble_eqn}
\ee
\be
{\ddot{R} \over R} \: = \:
-{4 \pi G_N \over 3 c^2} (\rho + 3 P - 2 \rho_\Lambda),
\label{acceleration}
\ee
where $H(t)$ is the Hubble expansion rate, the dark energy density is given by
$\rho_\Lambda \: = \: {\Lambda c^4 \over 8 \pi G_N}$,
$\rho$ represents the FL matter-energy density,
and $P$ is the pressure.
The term which involves the spatial curvature $k$ has
explicit scale dependence on the FRW parameter $R$.
The dark energy density is assumed to make a negligible contribution to
the FL expansion during de-coherence, but will become significant as the FL energy density
decreases due to the expansion of the universe.

The energy density during dark energy de-coherence $\rho_{DC}$ can be directly determined
from the Lemaitre equation to satisfy
\be
H_{DC} ^2 =
\left( {c \over R_{DC}} \right )^2 \: = \: {8 \pi G_N \over 3 c^2} \left ( \rho_{DC} + \rho_\Lambda \right ) -
{k c^2 \over R_{DC} ^2}.
\label{decoherence}
\ee
A so called
``open" universe ($k=-1$) is excluded from undergoing this transition, since
the positive dark energy density term $\rho_\Lambda$ already excludes a solution
with $\dot{R} \leq c$.
Likewise, for a ``closed" universe that is initially radiation dominated,
the scale factors corresponding to de-coherence $\dot{R}_{DC}=c$
and maximal expansion $\dot{R}_{max}=0$ can be directly compared.
From the Lemaitre equation
\be
{c^2 \over R_{max}^2} \: = \: { 8 \pi G_N \over 3 c^2} \left [ \rho (R_{max}) + \rho_\Lambda \right ] \: \cong \:
{ 8 \pi G_N \over 3 c^2} \rho_{DC} {R_{DC} ^4 \over R_{max} ^4} \Rightarrow
R_{max}^2 \: \cong \: 2 R_{DC} ^2 .
\ee
Clearly, this closed system never expands much beyond the transition scale.
Quite generally, the assumption of dark energy de-coherence due to sub-luminal
expansion at early times \emph{requires} that all such cosmologies are spatially flat.

Setting the expansion rate to $c$ in the Lemaitre equation \ref{Hubble_eqn} with $k=0$,
the energy density during dark energy de-coherence is given by
\be
\rho_{DC} \: = \:
{3 c^2 \over 8 \pi G_N}  \left ( {c  \over R_{DC} } \right )^2 \, - \, \rho_\Lambda.
\label{rhoFL}
\ee
Since the FL density at de-coherence is specified in terms of the single
parameter given by the scale at de-coherence $R_{DC}$, all results
which follow depend only on this single parameter.

As is often assumed, if the cosmology remains
radiation dominated in the standard way down to $t=0$, then the scale
parameter satisfies
\be
R(t) \: = \: R_{DC} \left ( {t \over t_{DC}} \right ) ^{1/2} ,
\ee
which gives the time at de-coherence as
\be
t_{DC} \: = \: {R_{DC} \over 2 c} .
\ee
The assumption of radiation dominance during de-coherence corresponds to a thermal
temperature of
\be
g(T_{DC}) \: (k_B T_{DC})^4 \: = \:
{90 \over 8 \pi^3} (M_P c^2)^2 \left ( {\hbar c \over R_{DC}} \right ) ^2 ,
\ee
where $g(T_{DC})$ counts the number of degrees of freedom associated
with particles of mass 
$m c^2 << k_B T_{DC}$, and $M_P=\sqrt{\hbar c/G_N}$ is the Planck mass.

Using the FL densities at radiation-matter (dust) equality $\rho_M (z_{eq})=\rho_{rad} (z_{eq})$,
one can extrapolate back to the de-coherence period to determine the redshift at that time.
Ignoring threshold effects (which give small corrections near particle thresholds while they
are non-relativistic), this gives
\be
1+z_{DC} \: = \: \left [ {\rho_{DC} \over \rho_{Mo}} (1+z_{eq})  \right ]^{1 \over 4} 
\: \cong  \: \left( {c \over H_o R_{DC}} \right )^{1 \over 2} \,
\left ( {1 + z_{eq} \over \Omega_{Mo}} \right ) ^{1 \over 4} ,
\label{zDC}
\ee
where equation \ref{rhoFL} has been used to explicitly exhibit the
$R_{DC}$ dependence.  Here, $\Omega_{Mo}$ is the present normalized mass density.
The scale parameter at the present time is then expressed in terms of this redshift
using the usual definition $R_o = (1+z_{DC}) R_{DC}$.

The evolution of the cosmology during the period
for which the dark energy density is constant and de-coupled 
from the FL energy density is
expected to be accurately modeled using the FL equations.  There is a period of
deceleration, followed by acceleration towards an approximately De Sitter expansion.
The rate of scale parameter expansion is sub-luminal during a
finite period of this evolution, as shown in Figure \ref{redshift}.
\twofigures{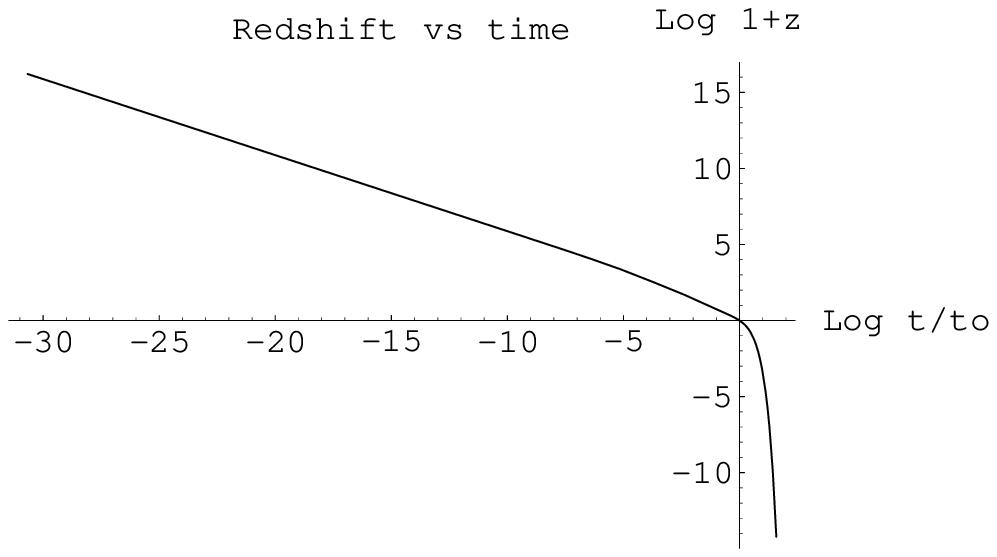}{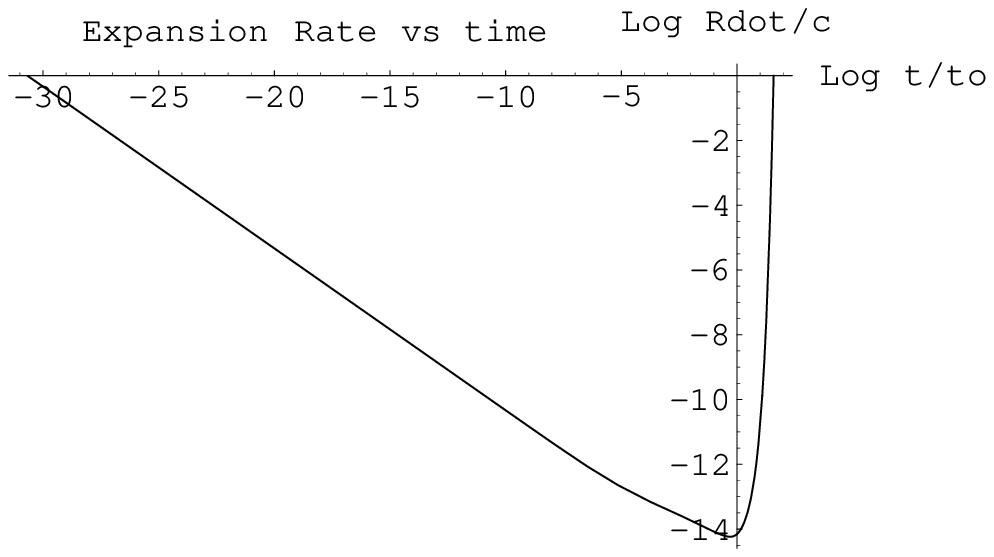}{Graphs of redshift and expansion rate vs time}
The particular value for the scale at de-coherence chosen for the graphs
is not important for the present discussion, and generally is determined by
the microscopic makeup of the dark energy.
The present time since the ``beginning" of the expansion
corresponds to the origin on both graphs.
The value of the expansion rate is by assumption equal to the speed of light for
any particular value chosen for $R_{DC}$, as well as when the expansion
scale reaches the De Sitter radius $R_\Lambda$.

Adiabatic perturbations are those that fractionally perturb the number
densities of photons and matter equally.  For adiabatic perturbations,
the matter density fluctuations grow according to\cite{PDG}
\be
\Delta \: = \: \left \{
\begin{array}{cc}
\Delta_{DC}  \left ( {R(t) \over R_{DC}} \right ) ^2  &
radiation-dominated \\
\Delta_{eq} \left ( {R(t) \over R_{eq}} \right ) &
matter-dominated
\end{array}
\right .
\ee
This allows an accurate estimation for the scale of fluctuations
at last scattering in terms of those during de-coherence
given by
\be
\Delta_{LS} \: = \: \left( {R_{LS} \over R_{eq}}  \right)
\left( {R_{eq} \over R_{DC}}  \right) ^2 \Delta_{DC} \: = \:
{ (1+z_{DC}) ^2 \over (1+z_{eq}) (1+ z_{LS}) } \Delta_{DC} .
\label{DelLS}
\ee

The energy available for fluctuations in the two point correlation
function is expected to be given by the cosmological dark energy,
in a manner similar to the background thermal energy $k_B T$
driving the fluctuations of thermal systems.  This means that the
amplitude of relative fluctuations $\delta \rho / \rho$
is expected to be of the order 
\be
\Delta_{DC} \: \equiv  \: \left (  {\rho_\Lambda \over \rho_{DC} }
\right ) ^{1/2} \: = \: {R_{DC} \over R_\Lambda}, 
\label{DelDC}
\ee 
where $\Lambda=3/R_\Lambda ^2$. Using equations
\ref{zDC}, \ref{DelLS}, and \ref{DelDC}, this amplitude at last scattering
is given by 
\be
\Delta_{LS} \: = \: { (1+z_{DC}) ^2 \over (1+z_{eq}) (1+ z_{LS}) }
 {R_{DC} \over R_\Lambda} \: \cong  \: {1 \over 1 + z_{LS}} \sqrt{{\Omega_{\Lambda o}
\over (1-\Omega_{\Lambda o}) (1+z_{eq})}} \cong 2.6 \times 10^{-5}, 
\ee 
where a spatially flat cosmology has been assumed.
This estimate is independent of the scale parameter during
de-coherence $R_{DC}$, and is of the order observed for
the fluctuations in the CMB (see \cite{PDG} section 23.2 page 221).
It is also in line with those argued by other
authors\cite{Padmanabhan}. Fluctuations in the CMB at last
scattering of this order are consistent with the currently
observed clustering of galaxies.

In conclusion, cosmological dark energy de-coherence has been assumed to occur when
the rate of expansion of the scale parameter
in the Friedmann-Lemaitre equations becomes sub-luminal.
The choice of a cosmological scale parameter in the FL equations
directly relates the scale of dark energy de-coherence
to the De Sitter scale associated with the cosmological constant.
Such a scale necessarily requires a spatially flat cosmology
in order to be consistent with structure formation.

When global gravitational coherence of the dark energy is lost, only local coherence
of microscopic degrees of freedom
within independent clusters is expected to remain, and the dark energy scale coherence
with the clusters is lost as the new degrees of freedom become available.  
The effect of dark energy density
at de-coherence is ``frozen out" as a positive cosmological constant.
In this interpretation, cosmological dark energy
need only be constant during the de-coherence epoch ($\dot{R} < c$).
Although de-coherence has been assumed to occur at
a particular value for the scale parameter $R_{DC}$, the
predicted order of magnitude for the amplitude of CMB
fluctuations has been shown to be independent of this scale
(and by inference, independent of the energy density) at de-coherence.

The dark energy de-coherence hypothesis defines a class of cosmological models all of which give
an amplitude of density fluctuations in the CMB expected to be of the order $10^{-5}$.
Work in progress involves specific models of the de-coherence process.
For these models,  the form of the power spectrum of the fluctuations expected to
be generated by de-coherence will be calculated.

\bigskip

We are grateful for useful discussions with E.D.Jones, W.R. Lamb, and T.W.B. Kibble.

\end{document}